\documentclass[onecolumn,superscriptaddress,showkeys,amsmath,amssymb,aps,prab,floatfix]{revtex4-2}
\usepackage{ulem}

\usepackage{graphicx}
\usepackage{color}
\usepackage{xcolor}
\usepackage{natbib}

\usepackage[utf8]{inputenc}
\usepackage{amsmath}
\usepackage{appendix}
\usepackage{graphicx}
\usepackage{siunitx}
\usepackage{amsthm,amstext}  
\usepackage{mathtools}           
\usepackage{physics} 
\usepackage{booktabs}
\usepackage{ulem}

\usepackage{xcolor}
\usepackage{adjustbox}
\usepackage{subcaption}
\usepackage{float}
\usepackage{upgreek}
\usepackage{rotating}
\usepackage{algorithm}
\usepackage{algpseudocode}

\newcommand{\sm}[1]{{\color{black}{#1}}} 
\newcommand{\ver}[1]{{\color{black}{#1}}}  

\usepackage{hyperref}
\hypersetup{
    colorlinks=true,
    citecolor=blue,
    linkcolor=blue,
    filecolor=blue,      
    urlcolor=blue,
}

\usepackage{subcaption}

\begin{document}

\title{Conversion of weighted macro-particle distributions}
\author{Nicolas Pichoff}
\address{CEA, Université Paris-Saclay, IRFU, DACM, Gif-sur-Yvette, France.}
\author{Samuel Marini}
\address{CEA, Université Paris-Saclay, IRFU, DACM, Gif-sur-Yvette, France.}
\date{\today}

\begin{abstract}
This article introduces a method for adjusting macro-particle weights within a particle distribution while preserving statistical and physical properties. The method allows the weights of the new macro-particle distribution to be determined by any continuous function, including uniform. Computational simulations validate the presented approach.
\end{abstract}

\maketitle
\section{Introduction}
The dynamics of particles moving in three-dimensional space can be represented by a six-dimensional vector, accounting for variations in both position and velocity. It requires, therefore, solving six differential equations for each particle, becoming computationally demanding whenever dealing with a large number of particles. A common solution in Particle-In-Cell (PIC) and Monte Carlo codes is to use macro-particles to represent real particles \citep{Derouillat2018,Lehe2016,Hara23,Uriot2015}. In these codes, each macro-particle has an associated weight designed to represent/sample a different number of real particles. The choice of weight for the macro-particles depends on the objective. For example, using macro-particles with a non-uniform distribution of weights is convenient to enhance the resolution in phase-space regions where the particle density is comparatively small. Nevertheless, the non-uniform distribution of weights reduces the statistical accuracy of the distribution properties. If $N$ is the number of macro-particles and $w_i$ their individual weights, the uncertainties on the evaluation of statistical and physical properties of the distribution is given by $\sqrt{\sum{w_i^2}/\left(\sum{w_i}\right)^2}$, being a minimum ($=1/\sqrt{N}$) only when all $w_i$ are the same. One can then define the statistical equivalent number of macro-particles as

\begin{equation}\label{eq1}
N_{eq} = \frac{\left(\sum{w_i}\right)^2}{\sum{w_i^2}} \leq N.
\end{equation}

Besides, some simulation codes operate only with macro-particles having uniform-weight, requiring, therefore, the conversion of non-uniform to uniform-weight macro-particle distributions \citep{Uriot2015}. In this regard, this article addresses a method for adjusting macro-particle weights within a distribution while preserving the distribution's key statistics and physical properties.

This article is organized as follows: Section II presents the mathematical support to validate the method for adjusting the macro-particle weights. Section III outlines the steps required to generate a macro-particle weight distribution, starting from an initial macro-particle weight distribution. Section IV presents results to illustrate the discussed methodology, and in Section V, we present our main conclusions.

\section{Mathematical development}

Let us consider an initial particle distribution $p$ with $n$ dimensions (here we assume $n = 6$ to represent a 6D phase-space) sampled by a set of $N$ weighted macro-particles of weight $w_i$ such that

\begin{equation}
p_{1\leq i \leq N}: (w_i,x_{i,1},...,x_{i,n}) = (w_i,\vec{x_i}),
\end{equation}
where
\begin{equation}
\vec{x_i} = \begin{pmatrix}
x_{i,1} \\
\vdots \\
x_{i,n}
\end{pmatrix}
\end{equation}
defines the phase-space coordinates of the macro-particles. The total weight of the distribution is then given by
\begin{equation}
W = \sum_{i=1}^{N} w_i,
\end{equation}
in which each macro-particle represents a fraction $w_i/W$ of the total distribution weigh.

Now, considering the transformed weighted statistical particle distribution $p'$ composed by $N'$ macro-particles of weight $w_j'$ such as
\begin{equation}
p'_{1\leq j \leq N'}: (w'_j,x'_{j,1},...,x'_{j,n}) = (w'_{j},\vec{x}'_{j}).
\end{equation}
For each macro-particle $i$ in the initial distribution, having weight $w_i$, a set of $v_i$ new macro-particles can be associated, each with its respective weights determined by a function $h$ so
\begin{equation}\label{eq6}
w'_{j} = h(w_{i}),
\end{equation}
in a way, the sum of the new weights is given by
\begin{equation}
W' = \sum_{j=1}^{N'} w'_{j}.
\end{equation}
We define the set $v_i$ as

\begin{equation}\label{eq8}
v_i = \frac{w_{i}/h(w_{i})}{\sum_{i=1}^{N} w_{i}/h(w_{i})} N',
\end{equation}
what implies $v_{i}=(w_i/W) N'$  when $w'_{j} = 1$ and  $v_{i}=N'/N$ when $w'_{j} = w_{i} $. Then, through the following algorithm, the value of $v_i$ associated with each macro-particle in the initial distribution allows us to determine how many new macro-particles should be generated in the new set, in such a way that

\begin{itemize}
    \item if $0 < v_i \leq 1$, the initial macro-particle represents less than 1 final macro-particle.
    \begin{itemize}
        \item Let $r$  be a number generated between 0 and 1 (with uniform probability).
        \subitem -If $r \leq v_i$, a new macro-particle following $p'_{j}: (w'_{j},\vec{x'}_{j})$ is generated and $j$ is incremented by $1$.
    \end{itemize}
    
    \item $v_i > 1$, the initial macro-particle represents more than 1 final macro-particle.
    \begin{itemize}
        \item Let $v_i = n_i + f_i$, where $n_i$ is an integer and $0 \leq f_i < 1$.
        \item Let $r$ be a number generated between 0 and 1 (with uniform probability).
        \subitem -If $r \leq f_i$, $n_i + 1$ new macro-particles following $p'_{j}: (w'_{j},\vec{x'}_{j})$ are generated
        \subitem -If $r > f_i$, $n_i$ new macro-particles following $p'_{j}: (w'_{j},\vec{x'}_{j})$ are generated.
    \end{itemize}
\end{itemize}

The subsequent discussion shows a methodology for generating the transformed weighted statistical particle distribution $p' (w'_{j},\vec{x'}_{j})$, focusing on the phase-space position of the generated macro-particles. 

\section{Generation of $\vec{x'}_{j}$ from associated $\vec{x}_{i}$}

Intuitively, one considers that the phase-space coordinate of the new set $\vec{x'}_{j}$ have to be close to the former phase-space coordinate $\vec{x}_{i}$, but not exact if one wants to avoid generating $ n_i+1 $ identical particles. We can then write $\vec{x'}_{j} = \vec{x}_{i} + \delta\vec{x'}_{j}$. On one side, the addition of the term $\delta\vec{x'}_{j}$ corresponds to a diffusion process. It has then to be as small as possible.
On the other side, if too small, the $n_i+1$ particles are quasi identical and they loss in statistical meaning since, statistically, identical particles contribute as one.

In order to balance these two aspects, we propose to adjust the range of $\delta\vec{x'}_{j}$ based on a characteristic distance between the macro-particles in a centered, normalized, and uncoupled frame. To achieve this transformation, a number of calculations are necessary and the details of these are presented below.

Initially, we transform the set $\vec{x}_i$ to a new set $\vec{X}_i$, which centers and normalizes the distribution such that:

\begin{equation} \label{eq9}
X_{i,k} = \frac{(x_{i,k}-\langle x_k \rangle)}{\sqrt{\sigma_{k,k}}},
\end{equation} 
where
\begin{equation}
\langle x_k \rangle = \sum_{i=1}^{N} \frac{w_i \cdot x_{i,k}}{W}
\end{equation} 
is the average of $x_k$ and $\sigma_{k,k}$ is the diagonal of a matrix that represents the second order momentum and gives the average sizes and correlation of the macro-particles along each dimension $k$ with $1 \leq k \leq n$, such as

\begin{equation}
\sigma = \begin{bmatrix}
\langle (x_1 - \langle x_1 \rangle) \cdot (x_1 - \langle x_1 \rangle) \rangle & \cdots & \langle (x_1 - \langle x_1 \rangle) \cdot (x_n - \langle x_n \rangle) \rangle \\
\vdots & \ddots & \vdots \\
\langle (x_n - \langle x_n \rangle) \cdot (x_1 - \langle x_1 \rangle) \rangle & \cdots & \langle (x_n - \langle x_n \rangle) \cdot (x_n - \langle x_n \rangle) \rangle
\end{bmatrix}.
\end{equation}

We propose then to transform $\vec{X}_i$ into a new set of coordinates $\vec{Y}_i$, in which, the beam is centered, normalized and also uncoupled, in a way that its corresponding sigma matrix is the identity matrix.
Let $[T]$ be the transfer matrix between $\vec{X}_i$ and $\vec{Y}_i$, such as

\begin{equation}\label{eq12}
\begin{bmatrix}
X_{1} \\
\vdots \\
X_{n}
\end{bmatrix} =
[T] \cdot
\begin{bmatrix}
Y_{1} \\
\vdots \\
Y_{n}
\end{bmatrix},
\end{equation}
Therefore, $[T]$ is a solution of equation:

\begin{equation}\label{eq13}
\begin{bmatrix}
1 & \cdots & \langle X_{1} \cdot X_{n} \rangle \\
\vdots & \ddots & \vdots \\
\langle X_{n} \cdot X_{1} \rangle & \cdots & 1
\end{bmatrix} =
[T] \cdot
\begin{bmatrix}
1 & 0 & \cdots & 0 \\
0 & 1 & \ddots & \vdots \\
\vdots & \ddots & \ddots & 0 \\
0 & \cdots & 0 & 1
\end{bmatrix} \cdot
[T]^{T}=[T]\cdot[T]^{T},
\end{equation}
where we can identify the matrix on the left side as the sigma matrix for the normalized and centralized set $\vec{X}_i$ and the identity matrix as the sigma matrix for the centered, normalized and also uncoupled set $\vec{Y}_i$. As Eq.~\eqref{eq13} implies many possible solutions for $[T]$ (36 free parameters with only 20 equations as $(X_{i}, X_{j}) = (X_{j}, X_{i}) $), we choose the solution in which $[T]$ is lower triangular, in a way that:

\begin{equation}
[T] \cdot [T]^T = 
\begin{bmatrix}
    T_{1,1} & 0 & \dots & 0 \\
    T_{2,1} & T_{2,2} & \dots & 0 \\
    \vdots & \vdots & \ddots & \vdots \\
    T_{n,1} & T_{n,2} & \dots & T_{n,n}
\end{bmatrix}
\cdot
\begin{bmatrix}
    T_{1,1} & T_{2,1} & \dots & T_{n,1} \\
    0 & T_{2,2} & \dots & T_{n,2} \\
    \vdots & \vdots & \ddots & \vdots \\
    0 & 0 & \dots & T_{n,n}
\end{bmatrix},
\end{equation}
and since $ T_{i,k} = 0 $ whenever $k > i$ (lower triangular shape), we can write \eqref{eq13} as
\begin{equation}
\sum_{k=1}^{\min(i,j)} T_{i,k} \cdot T_{j,k} = \langle X_i \cdot X_j \rangle.
\end{equation}

Considering $1 \leq i \leq n$, then $1 \leq j \leq i - 1$, the matrix $[T]$ reads (see demonstration in Appendix A):
\begin{equation}
T_{i,i} = \sqrt{\langle X^2_i \rangle - \sum_{k=1}^{i-1} T^2_{i,k}}
\end{equation}
and

\begin{equation}
T_{i,j} = \frac{\langle X_{i}X_{j} \rangle - \sum_{k=1}^{j-1} T_{i,k}T_{j,k}}{T_{j,j}}.
\end{equation}

Therefore, from \eqref{eq12}, we can write the centralized, normalized, and uncoupled distribution $\vec{Y}_i$ as:

\begin{equation}
\vec{Y}_{i} = [T]^{-1} \vec{X}_{i}.
\end{equation}

Given the distribution $\vec{Y}_i$, we can then create the new distribution $\vec{Y}_j'=\vec{Y}_i+\delta \vec{Y}_j$. To do this, we use the weight of each macro-particle in the initial distribution and project that weight (Eq. \eqref{eq6}) to calculate $v_i$ (Eq. \eqref{eq8}). Then, we apply the algorithm discussed at the end of Section II to decide whether a new macro-particle should be generated or not. The macro-particles of the new distribution are located at $\vec{Y}_i+\delta \vec{Y}_j$, with $\delta \vec{Y}_j$ given by a random Gaussian distribution of sigma size $\delta \sigma$, whose contribution to second order momentum should be small with respect to statistical uncertainties.
To estimate $\delta \sigma$, we start by defining an ideal Gaussian distribution with infinite particles, characterized by a mean of zero and a real standard deviation $\sigma_r$ of one. Introducing a random distribution composed of $N_{eq}$ particles from this ideal case incurs an inherent error, $\sigma_i = \sigma_r \pm \sigma_r/\sqrt{2N_{eq}}$. Generating a new distribution from a transformed one with $N_{eq}'$ particles introduces an additional error. Consequently, the final standard deviation can be expressed as $\sigma_f \approx \sigma_r \pm \sigma_r[ (1/\sqrt{2N_{eq}})^2 + (1/\sqrt{2N_{eq}'} )^2+ (\varphi/\sqrt{2N_{eq}'})^2]^{1/2}$, thereby incorporating two new terms attributable to the transformation and variation in particle position. This formulation allows us to define $\delta \sigma$ as $\delta \sigma \equiv \varphi \sigma_r/\sqrt{2N_{eq}'}$. The particle size variation term should remain minimal compared to the intrinsic error terms, which determines our choice for the parameter $\varphi$ as $\varphi \ll 1$. Moreover, it highlights the constant presence of intrinsic error within the distribution, which decreases as the number of particles used for its representation increases.

Note that at this stage, the choice of $\delta \sigma$ is subjective and can be optimized depending on the distribution and even, in a much longer way, depending on the position of each initial macro-particle ({\it e.g.}, by exploring the average distance between the neighboring macro-particles in the distribution).

Once we have generated $\vec{Y}_j'$, the final distribution is given by Eq. \eqref{eq9} and \eqref{eq12}, such that:

\begin{equation}
    p'_j: (w_j', x'_{j,1}...x'_{j,n})=(w_j',\vec{x}_j'),
\end{equation}
with  $x'_{j,k}=\langle x_k \rangle +\sqrt{\langle (x_k-\langle x_k \rangle )^2 \rangle}.([T].\vec{Y}_j')_k.$

\section{Results}

To demonstrate our methodology, we consider a basic 6D Gaussian uncoupled particle distribution, represented by $N$ macro-particles. We assume these particles have a uniform coordinate distribution $\vec{x}$ within $\pm 4$ sigmas, that is $-4\sigma_k \le x_{i=1,..N,k} \le 4\sigma_k$ with $k=1,...,n$. We have also assumed $\sigma_k=\sigma_{dist}=1$, $\varphi=0.01$, and the weight of each macro-particle is given by a 6D Gaussian distribution as follows

\begin{equation}\label{eq22}
w_i = \prod_{k=1}^{n=6}  e^{\left( - x_{i,k}^2/2\sigma_k^2 \right)} .
\end{equation} \label{eq.wi}

Moreover, our results presentation focuses on the centered-normalized and uncoupled distribution, both before and after applying the weight transformation, {\it i.e.,} the initial and final macro-particle positions are given by $\vec{Y}_i$ and $\vec{Y'}_j$. Therefore, in accordance with Eq.~\eqref{eq1}, the estimated number of equivalent particles is $N_{eq}^{i} \approx 8 \times 10^{-3} N$, where $N$ is the initial number of macro-particles. Assuming $N=1$M, then $N_{eq}^{i}=8000$ and the associated uncertainty of the initial distribution due to the non-homogeneity is given by $\sim 1/\sqrt{N_{eq}^{i}}=0.0111$. Effectively, we can verify this result by analyzing the difference between the sigma matrix of the initial distribution centralized and normalized $\vec{X}_i$, $\sigma_{\vec{X}_i}$, and the identity matrix $I$ such as

\begin{equation}
  \sigma_{\vec{X}_i}-I=
   \begin{bmatrix}
    \textcolor{red}{0} & 0.009870 & -0.001808 & -0.000034 & 0.001355 & -0.000489 \\
    0.009870 & \textcolor{red}{0} & -0.006048 & -0.002886 & 0.005133 & 0.001625 \\
    -0.001808 & -0.006048 & \textcolor{red}{0} & -0.005567 & 0.006023 & 0.005532 \\
    -0.000034 & -0.002886 & -0.005567 & \textcolor{red}{0} & 0.003009 & 0.006778 \\
    0.001355 & 0.005133 & 0.006023 & 0.003009 & \textcolor{red}{0} & 0.001287 \\
    -0.000489 & 0.001625 & 0.005532 & 0.006778 & 0.001287 & \textcolor{red}{0} \\
  \end{bmatrix},
\end{equation}

\noindent where the elements of the matrix $\sigma_{\vec{X}_i}-I$ indicates the uncertainty related to the non-homogeneity of the initial distribution and the value $0$ on the diagonal is due the normalisation of the macro-particle distribution.

A better insight is given by looking at the projection of the particle distribution. In this regard, Figs. 1 and 2 shows a 2D density projection, where the colors represent the density distribution on a Log$_{10}$ scale, spanning from $-2$ to $-8$ across six decades. In the figures, each pixel represents a step size of $0.1$ sigma.

Figure 1(a) shows the initial distribution of $N = 1$M macro-particles, where the initial weight is given by Eq.~\eqref{eq22} with $\sigma_k = 1 $. Figure 1(b) and (c) shows the final distribution with a uniform weight $w_j'= 1$, where the number of particles in the final distribution, denoted as $N'$, is $100$k in panel 1(b) with an uncertainty of $\sim1/\sqrt{N_{eq}}=0.00316$ and $10$k in panel 1(c) with an uncertainty of $\sim 1/\sqrt{N_{eq}}=0.01$.

\begin{figure}[H]
\centering
\includegraphics[width=5.5cm]{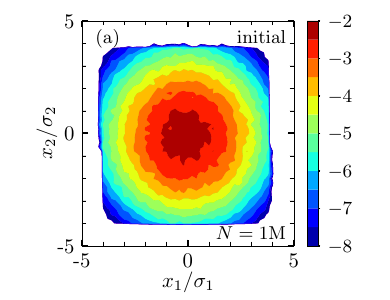}
\includegraphics[width=5.5cm]{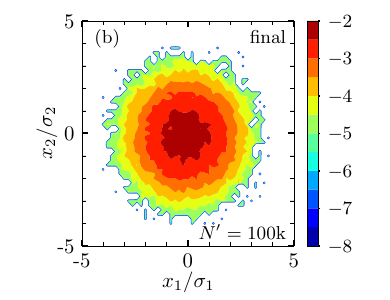}
\includegraphics[width=5.5cm]{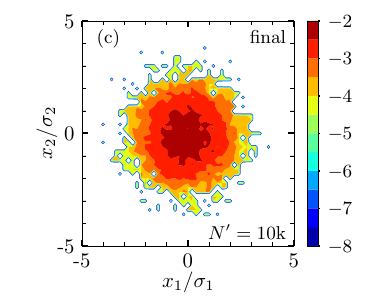}
\caption{2D macro-particle density distributions. Panel (a) displays the initial distribution with $N=1$M. Panels (b) and (c) exhibit the final distributions with $N^\prime=100$k and $N^\prime=10$k, respectively, after applying the weighting function $w_{j}' = 1$ constant.}
\label{fig:figure1}
\end{figure}

The centered-normalized sigma matrix for the particle distribution presented in Figure 1(a) and (b) is
\begin{equation}
  \sigma_{N'=100\text{k}}(w_{j}'= 1)-\sigma_{initial}=
\begin{bmatrix}
    \textcolor{red}{0.003000} & -0.002466 & 0.000639 & -0.000476 & -0.000626 & 0.001763 \\
    -0.002466 & \textcolor{red}{-0.002570} & 0.002477 & 0.001317 & 0.000167 & -0.000299 \\
    0.000639 & 0.002477 & \textcolor{red}{0.002779} & -0.001013 & -0.001071 & -0.000501 \\
    -0.000476 & 0.001317 & -0.001013 & \textcolor{red}{-0.002599} & 0.001357 & -0.001732 \\
    -0.000626 & 0.000167 & -0.001071 & 0.001357 & \textcolor{red}{-0.004562} & 0.002559 \\
    0.001763 & -0.000299 & -0.000501 & -0.001732 & 0.002559 & \textcolor{red}{0.001770} \\
  \end{bmatrix},
\end{equation}
and between Figure 1(a) and (c) is
\begin{equation}
 \sigma_{N'=10\text{k}}(w_{j}'= 1)-\sigma_{initial}=
\begin{bmatrix}
    \textcolor{red}{0.013471} & -0.003039 & -0.006285 & -0.003958 & -0.000749 & -0.005048 \\
    -0.003039 & \textcolor{red}{0.005882} & 0.011670 & 0.005178 & 0.006863 & 0.003610 \\
    -0.006285 & 0.011670 & \textcolor{red}{0.023162} & 0.003815 & -0.007525 & -0.009574 \\
    -0.003958 & 0.005178 & 0.003815 & \textcolor{red}{-0.017996} & -0.000913 & -0.005722 \\
    -0.000749 & 0.006863 & -0.007525 & -0.000913 & \textcolor{red}{-0.003248} & 0.006850 \\
    -0.005048 & 0.003610 & -0.009574 & -0.005722 & 0.006850 & \textcolor{red}{0.005470} \\
  \end{bmatrix}.
\end{equation}

The comparisons between the new and original centered-normalized sigma matrices for the particle distribution depicted in Figure 1 indicates small variations occurring with a reduced number of particles. Importantly, these variations generally do not modify the fundamental statistical and physical properties of the system. Indeed, the observed differences primarily highlight the minimal impact of statistical (shot) noise on the system. 
From a statistical standpoint, the difference between the initial and final sigma matrix represents the uncertainty in the distribution coefficients that is on the order of $\sim1/\sqrt{N_{eq}}$, as discussed.

Additionally, Figure 2 presents the initial distribution of $N = 1$M macro-particles with the initial weight given by Eq.~\eqref{eq22} with $\sigma_k = 1 $ and the final weight given by $ w_{j}' = \sqrt{w_i} $, considering that the number of macro-particles in the final distribution $N'$ is equal to $100$k in panel 2(b) and $10$k in panel 2(c).

\begin{figure}[H]
\centering
\includegraphics[width=5.5cm]{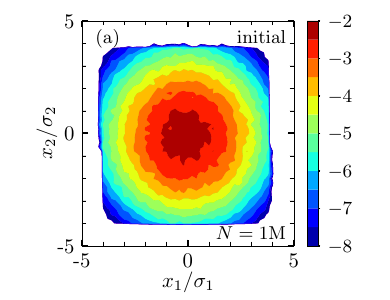}
\includegraphics[width=5.5cm]{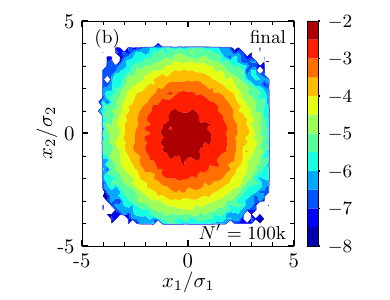}
\includegraphics[width=5.5cm]{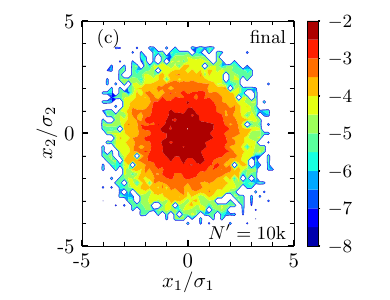}
\caption{2D macro-particle density distributions. Panel (a) displays the initial distribution with $N=1$M. Panels (b) and (c) exhibit the final distributions with $N^\prime=100$k and $N^\prime=10$k, respectively, after applying the weighting function $w_{j}' = \sqrt{w_i}$.}

\label{fig:figure2}
\end{figure}

Once more, these observed variations underscore the minimal influence of statistical noise. The centered-normalized sigma matrix of the particle distribution, as illustrated in Figures 2(a) and 2(b) is

\begin{equation}
  \sigma_{N'=100\text{k}}(w_{j}'=\sqrt{w_i})-\sigma_{initial}=
\begin{bmatrix}
    \textcolor{red}{-0.000777} & 0.000557 & 0.002895 & -0.000225 & -0.000179 & 0.000288 \\
    0.000557 & \textcolor{red}{0.000453} & 0.000514 & 0.000454 & -0.002519 & -0.001353 \\
    0.002895 & 0.000514 & \textcolor{red}{-0.000124} & -0.000720 & 0.001082 & -0.002994 \\
    -0.000225 & 0.000454 & -0.000720 & \textcolor{red}{0.000225} & -0.001001 & 0.001463 \\
    -0.000179 & -0.002519 & 0.001082 & -0.001001 & \textcolor{red}{0.000199} & 0.001292 \\
    0.000288 & -0.001353 & -0.002994 & 0.001463 & 0.001292 & \textcolor{red}{0.000669} \\
  \end{bmatrix},
\end{equation}

and between Figures 2(a) and 2(c) is
\begin{equation}
  \sigma_{N'=10\text{k}}(w_{j}'=\sqrt{w_i})-\sigma_{initial}=
\begin{bmatrix}
    \textcolor{red}{0.014927} & 0.010332 & 0.002222 & 0.011819 & -0.006087 & 0.007917 \\
    0.010332 & \textcolor{red}{0.003525} & -0.000283 & 0.009320 & -0.002940 & -0.003579 \\
    0.002222 & -0.000283 & \textcolor{red}{0.008191} & -0.021468 & -0.002384 & -0.023161 \\
    0.011819 & 0.009320 & -0.021468 & \textcolor{red}{0.021330} & -0.009624 & -0.005081 \\
    -0.006087 & -0.002940 & -0.002384 & -0.009624 & \textcolor{red}{-0.009184} & -0.001418 \\
    0.007917 & -0.003579 & -0.023161 & -0.005081 & -0.001418 & \textcolor{red}{-0.000679} \\
  \end{bmatrix}.
\end{equation}

\vspace{2cm}

\section{Conclusion}

This study introduced a method to transform macro-particle weights within a particle distribution, preserving its key statistical and physical properties. The effectiveness of this approach was demonstrated through computational simulations and, as discussed, it has a dependence on the typical distance between particles. Through a statistical discussion on the distribution's properties, we have introduced an approximate function to address this distance, indicating that an optimal solution might involve utilizing the typical distance between each particle's neighbors. This suggests that there is space for further improvement of the results, tailored to specific objectives. Moreover, the present transformation method enhances flexibility in particle distribution representation by allowing for any continuous weight function, including uniform weights. Importantly, it offers practical advantages in enhancing statistical accuracy and computational efficiency within particle simulation codes. Furthermore, this approach facilitates integration between codes operating with and without uniform weights.

\acknowledgments
The authors are grateful to DACM/CEA team for fruitful discussions.

\newpage

\appendix

\section{Step-by-step demonstration of $ T_{i,j} $}
Let,
\begin{equation}
T_{i,j} = \frac{(X_i \cdot X_j) - \sum_{k=1}^{j-1} T_{i,k} \cdot T_{j,k}}{T_{j,j}}
\end{equation}
so, if $ i = j = 1 $,

\begin{equation}
T_{1,1} = \sqrt{(X_{1})^2}.
\end{equation}
Now considering $ i = 1 $, $ 1 \leq j \leq n\text{D} $, it follows that

\begin{equation}
T_{j,1} = \frac{(X_{1} \cdot X_{j})}{T_{1,1}}.
\end{equation}
in a way, when $ i = j = 2 $,

\begin{equation}
T_{2,2} = \sqrt{(X_{2})^2 - T_{2,1}^2}.
\end{equation}
For the case $ i = 2 $, $ 3 \leq j \leq n\text{D} $, we deduce

\begin{equation}
\begin{split}
T_{3,1}T_{2,1} + T_{3,2}T_{2,2} &= (X_3 \cdot X_2) \\
T_{4,1}T_{2,1} + T_{4,2}T_{2,2} &= (X_4 \cdot X_2) \\
T_{j,1}T_{2,1} + T_{j,2}T_{2,2} &= (X_j \cdot X_2) \\
T_{j,2} &= \frac{(X_j \cdot X_2) - T_{j,1}T_{2,1}}{T_{2,2}},
\end{split}
\end{equation}
with $ i = j = 3 $, we have
\begin{equation}
T_{3,3} = \sqrt{(X_{3})^2 - T_{3,1}^2 - T_{3,2}^2}.
\end{equation}
Considering $ i = 3 $, $ 4 \leq j \leq n\text{D} $, we arrive at

\begin{equation}
\begin{split}
T_{4,1}T_{3,1} + T_{4,2}T_{3,2} + T_{4,3}T_{3,3} &= (X_4 \cdot X_3) \\
T_{5,1}T_{3,1} + T_{5,2}T_{3,2} + T_{5,3}T_{3,3} &= (X_5 \cdot X_3) \\
T_{j,1}T_{3,1} + T_{j,2}T_{3,2} + T_{j,3}T_{3,3} &= (X_j \cdot X_3) \\
T_{j,3} &= \frac{(X_j \cdot X_3) - T_{j,1}T_{3,1} - T_{j,2}T_{3,2}}{T_{3,3}}.
\end{split}
\end{equation}

\ldots

\noindent
Hence, for $ i = j $, one can derive
\begin{equation}
T_{i,i} = \sqrt{(X_{i})^2 - \sum_{k=1}^{i-1} T_{i,k}^2},
\end{equation}
and finally, for $ i + 1 \leq j \leq n\text{D} $, one obtains

\begin{equation}
T_{j,i} = \frac{(X_{i} \cdot X_{j}) - \sum_{k=1}^{i-1} T_{j,k} \cdot T_{i,k}}{T_{i,i}},
\end{equation}
where Eqs. (A8) and (A9) corresponds to Eqs. (16) and (17) of the present article.

\bibliographystyle{apsrev}
\bibliography{biblio}

\end{document}